\documentclass[conference,a4paper]{IEEEtran}

\usepackage{graphics} 
\usepackage{epsfig} 
\usepackage{mathptmx} 
\usepackage{times} 
\usepackage{amsmath} 
\usepackage{amssymb}  
\usepackage{stmaryrd}
\usepackage{color}

\usepackage{cite}

\ifCLASSINFOpdf
\else
\fi
\usepackage{dblfloatfix}
\usepackage{url}

\begin{document}
\title{The Velocity of the Decoding Wave for Spatially Coupled Codes on BMS Channels}

\author{\IEEEauthorblockN{Rafah El-Khatib and Nicolas Macris}
\IEEEauthorblockA{LTHC, EPFL, Lausanne, Switzerland\\
Emails: \{rafah.el-khatib,nicolas.macris\}@epfl.ch}}


%


\maketitle

\begin{abstract}
We consider the dynamics of belief propagation decoding of spatially coupled Low-Density Parity-Check codes. It 
has been conjectured that after a short transient phase,
the profile of ``error probabilities'' along the spatial direction of a spatially coupled code develops a uniquely-shaped
wavelike solution that propagates with constant velocity $v$. Under this assumption and for transmission over general Binary Memoryless Symmetric channels, we derive a formula for $v$. We also propose  approximations that are simpler to compute and support our findings using numerical data. 
\end{abstract}

\IEEEpeerreviewmaketitle

\section{Introduction}
Spatial coupling is a construction of Low-Density Parity-Check (LDPC) codes that has been shown to be capacity-achieving  on general Binary Memoryless Symmetric (BMS) channels under Belief Propagation (BP) decoding \cite{KRUUniv}.
The capacity-achieving property is due to the ``threshold saturation'' of the BP threshold of the coupled system towards the maximum a-posteriori (MAP) threshold of the 
uncoupled code ensemble \cite{KRUUniv}, \cite{kumar2014threshold}. 




To study the performance of codes under spatial coupling, it is useful to 
analyze the {\it decoding profile} along the spatial axis of coupling. For the sake of the discussion 
let the integer $z\in \{0, \dots, L-1\}$ denote 
the position along the spatial direction of the graph construction, where $L$ is the length of 
the coupling chain.
In the general framework of BMS channels 
the decoding profile consists of the vector of probability distributions of the log-likelihoods of 
bits under BP decoding. The $z$-th component of this vector, call it $\mathtt{x}_z$,
equals the log-likelihood {\it distribution} of the bits located at
the $z$-th position. In the special case of the Binary Erasure Channel (BEC) this reduces to a vector of erasure probabilities $0<x_z<1$.
This decoding profile satisfies a set of coupled Density Evolution (DE) iterative equations. It has been proven that under DE iterations,
as long as the channel
noise is below the MAP threshold, DE iterations drive $\mathtt{x}_z$ to the all-$\Delta_{\infty}$ vector (the Dirac mass at infinite log-likelihood, i.e. 
perfect knowledge of the bits) \cite{KRUUniv}, \cite{kumar2014threshold}. In the special case of the BEC 
this corresponds to a vector of erasure probabilities driven to zero
by DE iterations
\cite{yedla2014simple}. 


\begin{figure}
\centering
\includegraphics[scale=0.3]{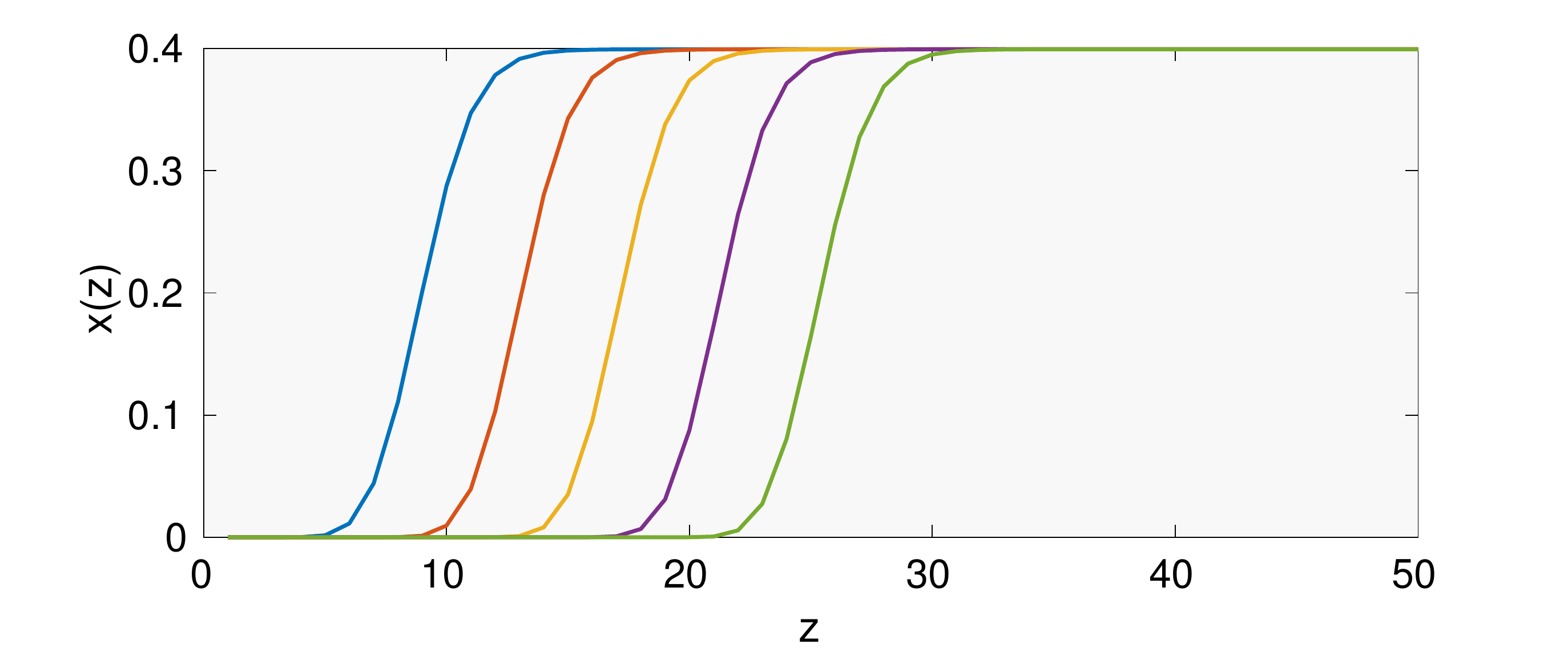}
\caption{We consider the $(3,6)$-regular LDPC spatially coupled code with $L=50$, $w=3$ on the BEC($0.46$). We plot the error probability along the spatial dimension and observe the ``decoding wave". This ``soliton" is plotted every 50 iterations until iteration 250 and is seen to make a quick transition from zero error probability to the BP-value of the error probability.}
\label{fig:wavePropagation}
\end{figure}

An interesting phenomenon that occurs during decoding, when the channel noise is between the BP and the MAP  
thresholds, is the appearance of a {\it decoding wave} or {\it soliton}. It has been observed
 that after a transient phase 
the decoding profile develops a {\it fixed shape} that seems independent of the 
initial condition and travels at 
{\it constant velocity} $v$. The soliton 
is depicted in Fig. \ref{fig:wavePropagation} for the case of a $(3,6)$-regular spatially coupled code on the BEC. 
Note that the 
same phenomenon has been observed for general BMS channels.


A solitonic wave solution has been proven to exist in \cite{KRU12} for the BEC, but the 
question of the independence of the shape from initial conditions is left open. In \cite{KRU12} and \cite{aref2013convergence},
bounds on the velocity of the wave for the BEC are proposed. In \cite{caltagirone2014dynamics} a formula for 
the wave in the context 
of the coupled Curie-Weiss toy model is derived and tested numerically.

In this work we derive a formula for the velocity of the wave in the {\it continuum limit} $L\gg w \gg 1$, with transmission over {\it general} BMS channels (see Equ. \eqref{eqn:vFormulaBMS}). Our derivation rests on the assumption that the soliton indeed appears. For simplicity we  limit ourselves to the case where the underlying uncoupled LDPC codes has only one nontrivial stable BP fixed point.

For the specific case of the BEC the formula greatly simplifies because density evolution reduces to a one-dimensional scalar system of equations. For general channels  we also apply the Gaussian approximation \cite{chung2001analysis}, which reduces the problem to a one-dimensional scalar system and yields a more tractable velocity formula.
We compare the numerical predictions of these velocity formulas 
with the empirical value of the velocity for finite $L$ and $w$. 

We propose a further approximation scheme (valid close to the MAP threshold) that expresses the velocity solely 
in terms of degree distributions of the code (in the spirit of \cite{KRU12}, \cite{aref2013convergence}). This may be useful to provide easy-to-handle design principles that maximize the velocity. 

Our formula can be applied to estimate parameters involved in the scaling law \cite{olmos2014scaling} of finite-size 
ensembles. This is briefly discussed as a further possible application.

The derivation of the velocity formula combines the use of  
the ``{\it potential functional}" introduced and used in a series of works (\cite{kumar2014threshold}, \cite{yedla2014simple}, \cite{hassani2012meanfield}, \cite{hassani2010itw})
and the continuum limit $L\gg w\gg 1$ which makes the derivations analytically tractable. 


In section \ref{section:setup}, we describe the setup and notation. In section \ref{section:formulaBMS}, we state our main result and show a sketch of the its derivation. The Gaussian approximation, the application to the BEC as well as further approximations, and the application to finite-size ensembles are discussed in section \ref{section:applicationsNumerics}. More details for some of the derivations can be found in \cite{elkhatibReport}.

\section{Potential formulation and continuum limit}\label{section:setup}
We consider (almost) the same setting and notation as in \cite{kumar2014threshold}. One important difference is that we will later consider the continuum 
limit, which is an approximation of the discrete expressions in the regime of large spatial length $L$ and 
window size $w$. 

\subsection{Preliminaries}\label{ssection:preliminaries}
Consider a symmetric probability measure $\mathtt{x}(\alpha)$ on the extended real numbers $\bar{\mathbb{R}}$.
These are measures satisfying 
$\mathtt{x}(\alpha)=e^\alpha \mathtt{x}(-\alpha)$ for all $\alpha\in\bar{\mathbb{R}}$. 
Here $\alpha\in\bar{\mathbb{R}}$ is interpreted as a ``log-likelihood variable". 
The (linear) entropy functional is 
defined as 
\begin{align}
H(\mathtt{x}) = \int\mathrm{d}\mathtt{x}(\alpha)\log_2(1+e^{-\alpha})
\end{align}
and will play an important role.

In the sequel we will use the Dirac masses $\Delta_0(\alpha)$ and $\Delta_\infty(\alpha)$ at zero and infinite likelihood, respectively. We will also need the standard variable-node and check-node convolution operators $\varoast$ and $\boxast$ for log-likelihood ratio (LLR) message distributions involved in DE equations (see \cite{richardson2008modern}). 
%

\subsection{Single system}\label{ssection:singleSystem}
Consider an LDPC($\lambda,\rho$) code ensemble and transmission over the BMS channel. 
Here $\lambda(y)=\sum_{l} \lambda_l y^{l-1}$ and $\rho(y)=\sum_{r} \rho_r y^{r-1}$ are the usual edge-perspective variable-node and check-node degree distributions. The node-perspective degree distributions $L$ and $R$ are 
defined by $L'(y)=L'(1)\lambda(y)$ and $R'(y)=R'(1)\rho(y)$.

We denote by $\mathtt{x}^{(t)}(\alpha)$ the variable-node output distribution at time $t\in\mathbb{N}$. 
We consider a family of BMS channels whose distribution $\mathtt{c}_\mathtt{h}(\alpha)$ in the log-likelihood domain is parametrized by the 
channel entropy $H(\mathtt{c}_\mathtt{h})=\mathtt{h}$. On the BEC, 
for example, we have $\mathtt{c}_\epsilon(\alpha)=\epsilon\Delta_0(\alpha)+(1-\epsilon)\Delta_\infty(\alpha)$, $H(\mathtt{c}_\epsilon)=\epsilon$.

We can track the average behavior of the BP decoder by means of the DE iterative equation \cite{richardson2008modern} \begin{align}
\mathtt{x}^{(t+1)}=\mathtt{c}_\mathtt{h}\varoast\lambda^\varoast(\rho^\boxast(\mathtt{x}^{(t)}))
\end{align}
with initial condition $\mathtt{x}^{(0)}=\Delta_0$.
The BP threshold $\mathtt{h}_{\text{\tiny BP}}$ is the largest value of $\mathtt{h}$ for which the DE recursion converges to $\Delta_\infty$.  

From now on we will omit the subscript $\mathtt{h}$ and the argument $\alpha$ of the distribution $\mathtt{c}_\mathtt{h}(\alpha)$ to alleviate notation. Later on, $\mathtt{c}_z$ and $\mathtt{c}(z)$ describe the channel distribution at the {\it spatial position} $z$ in {\it discrete} and {\it continuous} settings, respectively.



The potential functional $W_s(\mathtt{x};\mathtt{c})$ (of the ``single'' or uncoupled system) is
\begin{align}
W_s(\mathtt{x};\mathtt{c}) & =\frac{1}{R'(1)}H(R^\boxast(\mathtt{x}))+H(\rho^\boxast(\mathtt{x}))
\nonumber \\
&-H(\mathtt{x}\boxast\rho^\boxast(\mathtt{x}))-\frac{1}{L'(1)} H(\mathtt{c}\varoast L^\varoast(\rho^\boxast(\mathtt{x}))).
\label{eqn:potentialSingleBMS}
\end{align}
The DE equation is obtained by setting to zero the functional derivative of $W_s(\mathtt{x};\mathtt{c})$ with respect to $\mathtt{x}$.

The BP threshold $\mathtt{h}_{\text{\tiny BP}}$ is strictly smaller than the MAP threshold $\mathtt{h}_{\text{\tiny MAP}}$. 
Spatial coupling, however, exhibits the attractive property of {\it threshold saturation} which makes it possible to
decode perfectly up till $\mathtt{h}_{\text{\tiny MAP}}$.
The definitions of the BP and MAP thresholds above extend to the spatially coupled setting.
 
\subsection{Spatially coupled system}\label{ssection:spatiallyCoupledSystem}
Since the natural setting for coupling is discrete, we first 
describe the system in discrete space before taking the continuum limit.

The coupled LDPC($\lambda,\rho$) code ensemble is defined as follows. Consider $L+w$ ``replicas" of the single 
system described in section \ref{ssection:singleSystem}, on the spatial coordinates $z\in\{-w+1,...,L\}$. The system 
at position $z$ is coupled to other systems by means of a uniform coupling window
of width $w$. 
We denote by $\mathtt{x}_z^{(t)}$ the check-node input distribution at position $z\in\{-w+1,...,L\}$ on the spatial axis, 
and at time $t\in\mathbb{N}$. We then write the DE equation of the coupled system in discrete space as 
\begin{align}
\mathtt{x}_z^{(t+1)}=\frac{1}{w}\sum\limits_{i=0}^{w-1}\mathtt{c}_{z-i}
\varoast\lambda^\varoast\Bigg(\frac{1}{w}\sum\limits_{j=0}^{w-1}\rho^\boxast\Big(\mathtt{x}_{z-i+j}^{(t)} \Big) \Bigg).
\label{eqn:DEcoupledDiscrete}
\end{align}
Here $\mathtt{c}_z=\mathtt{c}$, for $z\in\{0,\dots, L\}$ and $\mathtt{c}_z=\Delta_\infty$ otherwise. We fix the left boundary to $\mathtt{x}_z^{(t)}=\Delta_\infty$ for $z\in\{-w+1,...,-1\}$, for all $t\in\mathbb{N}$. The initial condition on the right side is $\mathtt{x}_z^{(0)}=\Delta_0$, 
for $z\in\{0,...,L\}$.
The initialization to perfect 
information at the left boundary is what allows seed propagation along the chain of coupled codes. 

We denote by $\underline{\mathtt{x}}$ the profile vector. Then the expression of the discrete potential functional is
\begin{align}
&U(\underline{\mathtt{x}};\mathtt{c})=\sum\limits_{z=-w+1}^{L}\Bigg\{\frac{1}{R'(1)}H(R^\boxast(\mathtt{x}_z))+H(\rho^\boxast(\mathtt{x}_z)) \label{coupledpotentialdiscrete} \\
&-H(\mathtt{x}_z\boxast\rho^\boxast(\mathtt{x}_z))-\frac{1}{L'(1)} H\Big(\mathtt{c}_z\varoast L^\varoast\Big(\frac{1}{w}\sum\limits_{i=0}^{w-1}\rho^\boxast(\mathtt{x}_{z+i})\Big)\Big)\Bigg\}.\nonumber
\end{align}
The coupled DE equation \eqref{eqn:DEcoupledDiscrete} is obtained by setting to zero the functional derivative of this potential with respect to $\underline{\mathtt{x}}$.

\subsection{Continuum Limit}\label{ssection:contLimit}
We now consider the coupled system in the {\it continuum limit} 
(see \cite{KRU12}, \cite{el2013displacement}, \cite{el2014analysis} for the case of the BEC)
$L\rightarrow +\infty$ and then $w\rightarrow +\infty$. 
We set $\mathtt{x}(\frac{z}{w},t)\equiv \mathtt{x}_z^{(t)}$ and replace $\frac{z}{w}\to z$ 
where the new $z$ is a {\it continuous} variable on the spatial axis, $z\in \mathbb{R}$
(we slightly abuse notation here).
The DE equation \eqref{eqn:DEcoupledDiscrete} becomes
\begin{align*}
\mathtt{x}(z,t+1)=\int_0^1\mathrm{d}u\,\mathtt{c}(z-u)\varoast\lambda^\varoast\Big(\int_0^1\mathrm{d}s\,\rho^\boxast\big(\mathtt{x}(z-u+s,t) \big) \Big),
\end{align*}
where $\mathtt{c}(z)$ is now the BMS channel distribution at the continuous spatial position $z\in \mathbb{R}$ (we again slightly abuse notation) and the 
boundary / initial conditions are  $\mathtt{x}(z,t)=\Delta_\infty$, for $z<0$ and all $t\in\mathbb{R}$ / $\mathtt{x}(z,0)=\Delta_0$ for $z\geq 0$.

%
The potential functional $\mathcal{W}(\mathtt{x};\mathtt{c})$ of the coupled system in the continuum limit is obtained from 
\eqref{coupledpotentialdiscrete}. In order to get a finite result when $L\to +\infty$ we must normalize the potential by subtracting an ``energy" associated to a fixed profile $\mathtt{x}_{0}$ that satisfies $\mathtt{x}_{0}(z,t)\to\Delta_\infty$ when $z\to-\infty$ and $\mathtt{x}_{0}(z,t)\to\mathtt{x}_{\text{\tiny BP}}$ when $z\to+\infty$ ($\mathtt{x}_{\text{\tiny BP}}$ is the nontrivial stable fixed point log-likelihood density of BP for the single uncoupled code ensemble), for all $t\in\mathbb{N}$. The functional is thus defined as follows, 
\begin{align*}
\mathcal{W} & =
\int_{\mathbb{R}}\mathrm{d}z\,\Big\{\frac{1}{R'(1)}\Big(H(R^\boxast(\mathtt{x}(z,t)))-H(R^\boxast(\mathtt{x}_{0}(z,t)))\Big)
\nonumber \\
&\qquad +H(\rho^\boxast(\mathtt{x}(z,t)))-H(\mathtt{x}(z,t)\boxast\rho^\boxast(\mathtt{x}(z,t)))\nonumber\\
&\qquad -H(\rho^\boxast(\mathtt{x}_{0}(z,t)))+H(\mathtt{x}_{0}(z,t)\boxast\rho^\boxast(\mathtt{x}_{0}(z,t)))\nonumber\\
&\qquad -\frac{1}{L'(1)}H\Big(\mathtt{c}(z)\varoast L^\varoast\big(\int_0^1\mathrm{d}s\,\rho^\boxast(\mathtt{x}(z+s,t)) \big) \Big)\nonumber\\
&\qquad +\frac{1}{L'(1)}H\Big(\mathtt{c}(z)\varoast L^\varoast\big(\int_0^1\mathrm{d}s\,\rho^\boxast(\mathtt{x}_{0}(z+s,t)) \big) \Big)\Big\}.\nonumber
\end{align*}
It can be shown that the integral converges
under suitable assumptions on the entropy of $\mathtt{x}$ as $z\to\pm \infty$.

Once the functional 
derivative of $\mathcal{W}$ {\it in the direction} $\eta$ is computed, one finds that the DE equation is equivalent to
\begin{align}
\int_{\mathbb{R}}\mathrm{d}z\,H\Big((\mathtt{x}(z, t+1) - & \mathtt{x}(z,t))
\varoast[\rho^{\prime\boxast}(\mathtt{x}(z,t))\boxast \eta(z,t)] \Big)
\nonumber \\ &
=\frac{\delta \mathcal{W}}{\delta \mathtt{x}}[\eta(z,t)]
\label{eqn:FtnalDeriv}
\end{align}
This is a gradient descent equation in an infinite-dimensional space of measures.

\section{Main Result}
\label{section:formulaBMS}
\subsection{Velocity Formula}
We restrict ourselves to code ensembles with a single nontrivial stable BP fixed point. 
Consider the case when the channel entropy $\mathtt{h}$ satisfies $\mathtt{h}_{\text{\tiny BP}} < \mathtt{h} < \mathtt{h}_{\text{\tiny MAP}}$.
After a few iterations of DE, which we call the ``transient phase", one observes a solitonic (wavelike) behavior as depicted in Figure \ref{fig:wavePropagation}. 
This motivates us to make the following assumptions: (i) after a transient phase the profile develops a fixed shape
$\mathtt{X}(\cdot)$; 
(ii) the shape is independent of the initial condition; (iii) the shape travels at constant speed $v$; (iv) the shape satisfies the boundary 
conditions $\mathtt{X}(z)\to \Delta_{\infty}$ for $z\to -\infty$ and $\mathtt{X}(z)\to \mathtt{x}_{\text{\tiny BP}}$ for $z\to +\infty$. We thus make the ansatz $\mathtt{x}(z,t)=\mathtt{X}(z-vt)$.
\begin{figure}
\centering
\includegraphics[scale=0.6]{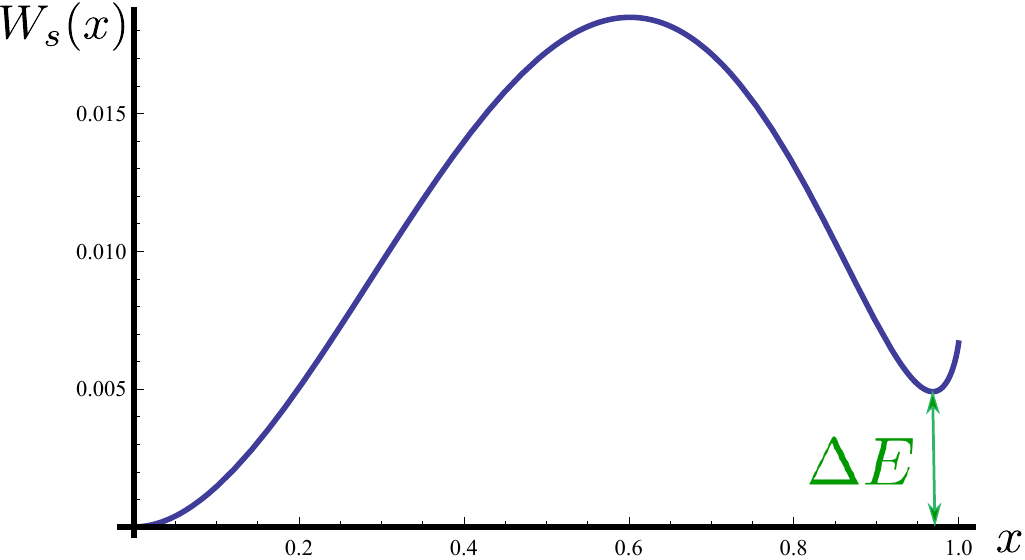}
\caption{The energy gap $\Delta E$ is the difference in the single potential at its stable fixed points. 
We plot the potential of the uncoupled $(3,4)$-regular code 
ensemble at $\epsilon=0.73$. Here $\epsilon_{\text{\tiny BP}}= 0.6473$ and $\epsilon_{\text{\tiny MAP}}=0.7456$. 
The gap always vanishes when $\epsilon=\epsilon_{\text{\tiny MAP}}$.}
\label{fig:energyGap}
\end{figure}

From this ansatz and equation \eqref{eqn:FtnalDeriv} we obtain our {\it main result}. The {\it velocity of the soliton} is
(primes are derivatives)
\begin{align}\label{eqn:vFormulaBMS}
v=\frac{\Delta E}{\int_{\mathbb{R}}\mathrm{d}z\,H\Big(\rho^{\prime\boxast}(\mathtt{X}(z))\boxast(\mathtt{X}'(z))^{\boxast 2}\Big)},
\end{align}
where $\Delta E$ is the {\it energy gap} defined as 
$\Delta E=W_s(\mathtt{x}_{\text{\tiny BP}};\mathtt{c})-W_s(\Delta_\infty;\mathtt{c})$.
We recall $W_s(\cdot;\cdot)$ is the potential of the {\it uncoupled system} \eqref{eqn:potentialSingleBMS}, and $\Delta_\infty$ the trivial fixed point (Dirac mass at infinity).
With our normalizations $W_s(\Delta_{\infty}; c)=0$.
Fig. \ref{fig:energyGap} illustrates the energy gap $\Delta E$ for the $(3,4)$-regular code ensemble on the BEC($\epsilon$). 

Formula \eqref{eqn:vFormulaBMS} involves the soliton shape $\mathtt{X}$. Using the DE equation and $x(z,t)= X(z-vt)$ we find that $\mathtt{X}(z)$ is the solution of 
\begin{align}
\mathtt{X}(z)&-v\mathtt{X}'(z)
=\int_0^1\mathrm{d}u\,\mathtt{c}\varoast\lambda^\varoast\Big(\int_0^1\mathrm{d}s\,\rho^\boxast\Big(\mathtt{X}(z-u+s) \Big) \Big).\label{eqn:diffEqnXShape}
\end{align}
Equs. \eqref{eqn:vFormulaBMS}-\eqref{eqn:diffEqnXShape} form a closed system of equations that can be solved iteratively to obtain $\mathtt{X}$ and $v$.

\subsection{Brief Sketch of Derivation}\label{ssection:formulaBMSderivation}

Consider equation \eqref{eqn:FtnalDeriv}. Under 
the ansatz $\mathtt{x}(z,t)=\mathtt{X}(z-vt)$ for small $v$ we get $\mathtt{x}(z, t+1) - \mathtt{x}(z,t)\approx -v\mathtt{X}'(z-vt)$.
Choosing the direction $\eta(z,t)=\mathtt{X}'(z-vt)$  we can rewrite (after a few manipulations involving properties of 
$\otimes$ and $\boxplus$) the left-hand side of \eqref{eqn:FtnalDeriv} as
\begin{align*}
v\int_{\mathbb{R}}\mathrm{d}z\,H\Big(\rho^{\prime\boxast}(\mathtt{X}(z))\boxast \mathtt{X}'(z)^{\boxast 2} \Big).
\end{align*}
We now consider the right-hand side of \eqref{eqn:FtnalDeriv}, which is the functional 
derivative of $\mathcal{W}(\mathtt{x};\mathtt{c})$ in the direction of $\eta(z,t)=\mathtt{X}'(z-vt)$. 
We first split $\mathcal{W}$ into two parts: the single system potential $\mathcal{W}_s(\mathtt{x};\mathtt{c})$ 
that remains if we ignore the coupling effect, and the rest which constitutes the ``interaction potential" 
$\mathcal{W}_i(\mathtt{x};\mathtt{c})$ (see \cite{el2013displacement} for similar splittings or \cite{elkhatibReport} for the exact definitions).
With some care one can then show that 
\begin{align*}
\frac{\delta \mathcal{W}_s}{\delta \mathtt{x}}[\mathtt{X}'(z)]=W_s(\mathtt{x}_{\text{\tiny BP}};\mathtt{c})-W_s(\Delta_\infty;\mathtt{c}),\quad\frac{\delta \mathcal{W}_i}{\delta \mathtt{x}}[\mathtt{X}'(z)]=0.
\end{align*}
We conclude that the ``interaction" part does not contribute to the velocity and only the energy gap remains. 

\section{Applications}\label{section:applicationsNumerics}
\subsection{Gaussian Approximation}
\label{section:gaussApprox}
In order to simplify the analysis of a spatially coupled ensemble with transmission over general BMS channels, one may use the method of Gaussian approximations \cite{chung2001analysis}. The idea is to assume that the densities of the LLR messages appearing in the DE equations are symmetric Gaussian densities (symmetric Gaussian densities 
$\mathtt{x}(\alpha)=(\sqrt{2\pi}\sigma)^{-1} \exp(-(\alpha -m)^2/2\sigma^2)$ 
are characterized by the relation $\sigma^2 = 2m$).  
Furthermore the channel density $\mathtt{c}$ is replaced by a BIAWGNC($\sigma_n^2$)
with the same entropy $H(\mathtt{c})$. 

The system becomes scalar one-dimensional since one only tracks the evolution of the means or equivalently the 
entropies of the densities. 
For simplicity, we consider the $(\ell,r)$-regular ensemble. Density evolution is conveniently expressed
in terms of the entropies $p(z,t)=H(\mathtt{x}(z, t))$. One then
observes a ``scalar" wave propagation much like the one of Fig. \ref{fig:wavePropagation}. 
The entropy of a symmetric Gaussian density of mean $m$
equals 
\begin{align*}
\psi(m)=\big(\sqrt{4\pi m}\,\big)^{-1}\int_{\mathbb{R}}\mathrm{d}z\,e^{-(z-m)^2/4m}\log_2(1+e^{-z}).
\end{align*}
With this function, the {\it Gaussian approximation for the velocity} reads (primes are derivatives)
\begin{align}
v_{\text{\tiny GA}}=\frac{W_s^{\text{\tiny GA}}(p_{\text{\tiny BP}};c)-W_s^{\text{ \tiny GA}}(0;c)}{-(r-1)\int_\mathbb{R}\mathrm{d}z\,
(p^\prime(z))^2\frac{\psi''((r-2)\psi^{-1}(1-p(z)))}{(\psi'(\psi^{-1}(1-p(z))))^2}},\label{eqn:velocityGauss}
\end{align}
where $p(z)$ denotes the shape the entropy profile, $p_{\text{\tiny BP}}=H(\mathtt{x}_{\text{\tiny BP}})$, 
and \eqref{eqn:potentialSingleBMS} now becomes
\begin{align*}
&W_s^{\text{\tiny GA}}(p; c)=\big(1-\frac{1}{r}\big)\psi\big(r\psi^{-1}(1-p)\big)-\psi\big((r-1)\psi^{-1}(1-p)\big)\\
&+\frac{1}{r}-\frac{1}{\ell}\psi\big(\psi^{-1}(H(\mathtt{c}))+\ell\psi^{-1}\big(1-\psi((r-1)\psi^{-1}(1-p)) \big) \big)
\end{align*}
(here $\psi^{-1}(H(\mathtt{c}))$ is the mean of the BIAWGNC($\sigma_n$) that has the same entropy as the channel $\mathtt{c}$ and equals $2/\sigma_n^2$). The shape $p(z)$ is computed from
\begin{align*}
&p(z) - v_{\text{\tiny GA}}p^\prime(z) = 
\int_0^1\mathrm{d}u\, \psi\Big(\psi^{-1}(H(\mathtt{c}))
\\ &
+ (\ell-1) \psi^{-1}\Big(1-\int_0^1 \mathrm{d}s\, \psi\big((r-1)\psi^{-1}(1-p(z-u+s))\big)\Big)\Big).
\end{align*}

\subsection{Velocity on the BEC}\label{ssection:formulaBECstatement}
For the BEC we can directly simplify  \eqref{eqn:vFormulaBMS}
(alternatively one can rederive the formula using directly the continuum approximation over the BEC). 
For this case, the channel distribution is 
$\mathtt{c} = \epsilon\Delta_0 + (1-\epsilon)\Delta_{\infty}$.
The fixed shape of the decoding wave is entirely characterized by the (scalar) erasure probability $x(z)$, i.e., 
$ \mathtt{X}(z)=x(z)\Delta_0 + (1-x(z))\Delta_{\infty}$ and $\mathtt{X}^\prime(z)=x'(z)\Delta_0-x'(z)\Delta_\infty$. Then 
using $\mathtt{X}(z)\varoast\Delta_0=\mathtt{X}(z)$, $\mathtt{X}(z)\varoast\Delta_\infty=\Delta_\infty$, $\mathtt{X}(z)\boxast\Delta_0=\Delta_0$, 
$\mathtt{X}(z)\boxast\Delta_\infty=\mathtt{X}(z)$, we find 
$\lambda^{\varoast}(\mathtt{X}(z)) \to \lambda(x(z))$, $\rho^{\boxast}(\mathtt{X}(z)) \to 1-\rho(1-x(z))$. 
The velocity becomes 
\begin{align}
v_{\text{\tiny BEC}}=\frac{W_{\text{\tiny BEC}}(x_{\text{\tiny BP}};\epsilon)-W_{\text{\tiny BEC}}(0;\epsilon)}{\int_{\mathbb{R}}\mathrm{d}z\,\rho^\prime(1-x(z))(x^\prime(z))^2},\label{eqn:formulaBEC}
\end{align}
where the single potential \eqref{eqn:potentialSingleBMS} now is  
\begin{align*}
W_{\text{\tiny BEC}}(x;\epsilon)=\frac{1-R(1-x)}{R'(1)}-x\rho(1-x)-\frac{\epsilon L(1-\rho(1-x))}{L'(1)} .
\end{align*}
Again, the erasure profile has to be computed from the one-dimensional equation 
$$
x(z) - v_{\text{\tiny BEC}}x^\prime(z) = \epsilon\int_0^1 \mathrm{d}u\, \lambda\Big(\int_0^1 \mathrm{d}s\, (1- \rho(1 - x(z-u+s))\Big).
$$

The formula obtained here for the BEC is obviously very similar to the {\it upper bound} proved in 
\cite{aref2013convergence} (Theorem 1) for a discrete system
\begin{align}
v_{\text{\tiny B}}=\alpha\frac{W_{\text{\tiny BEC}}(x_{\text{\tiny BP}};\epsilon) - W_{\text{\tiny BEC}}(0;\epsilon)}{\sum\limits_{z\in\mathbb{Z}}\rho^\prime(1-x_z)(x_z-x_{z-1})^2}, \qquad \alpha \leq 2.
\end{align}
In \cite{aref2013convergence} it is conjectured based on numerical simulations that 
$\alpha=1$ would be a tight bound. Our results here are perfectly consistent with the findings of \cite{aref2013convergence}. 

\subsection{Further approximations for scalar systems}
The one-dimensional formulas for the velocity might help design degree distributions. However it is costly to compute the shape of the profile (that enters the denominators) for every degree distribution.
We propose a hierarchy of approximations 
for the velocity that involve only the degree distributions  and quantities related to the single system, i.e, no profile shape 
needs to be computed. These are good for $\epsilon$ close to $\epsilon_{\text{\tiny MAP}}$. The first two approximations of the hierarchy are, for $i=1,2$,
\begin{align}
v_{a_i}&=\frac{W_{\text{\tiny BEC}}(x_{\text{\tiny BP}};\epsilon)-W_{\text{\tiny BEC}}(0;\epsilon)}{\int_0^{x_{\text{\tiny BP}}}\mathrm{d}x\,\rho^\prime(1-x)\sqrt{-12x^2+24F_i(x)}}
\end{align}
where $F_i(x)=\int_0^{x}\mathrm{d}y\,(1-\rho^{-1}(1-\lambda^{-1}(f_i(y)/\epsilon)))$ and 
where $f_1(y)=y$, $f_2(y)=y-v_{a_1}\sqrt{-12y^2+24F_1(y)}$.
The derivation is not shown here due to length constraints, but 
is to some extent inspired from that of \cite{caltagirone2014dynamics} for the coupled Curie-Weiss model introduced in \cite{hassani2012meanfield}, \cite{hassani2010itw}. We note that the approximation scheme breaks down if the quantities under the square roots are negative; this depends on the code parameters.

One can also derive similar approximative formulas within the Gaussian approximation which also is a 
one-dimensional scalar system. This is not shown here.

\subsection{Numerical Simulations}\label{ssection:simulations}
In this section we compare numerical predictions with the velocity formulas for the cases of the BEC($\epsilon$) and 
BIAWGNC($\sigma_n^2$), to the further approximations, and also to the empirical velocity.  
%
%
The empirical velocity $v_e$ is the velocity calculated from the decoding profiles obtained by running DE \eqref{eqn:DEcoupledDiscrete}. In particular, it is equal to the average of $\Delta z/(w\Delta I)$, where $\Delta z$ is the spatial distance between the centers of the kinks (or fronts) of two profiles, and $\Delta I$ is the number of iterations of DE that were made to go from the first profile to the second. It serves as the reference value of the velocity with which we compare our formula.

\begin{figure}
\centering
\includegraphics[scale=0.27]{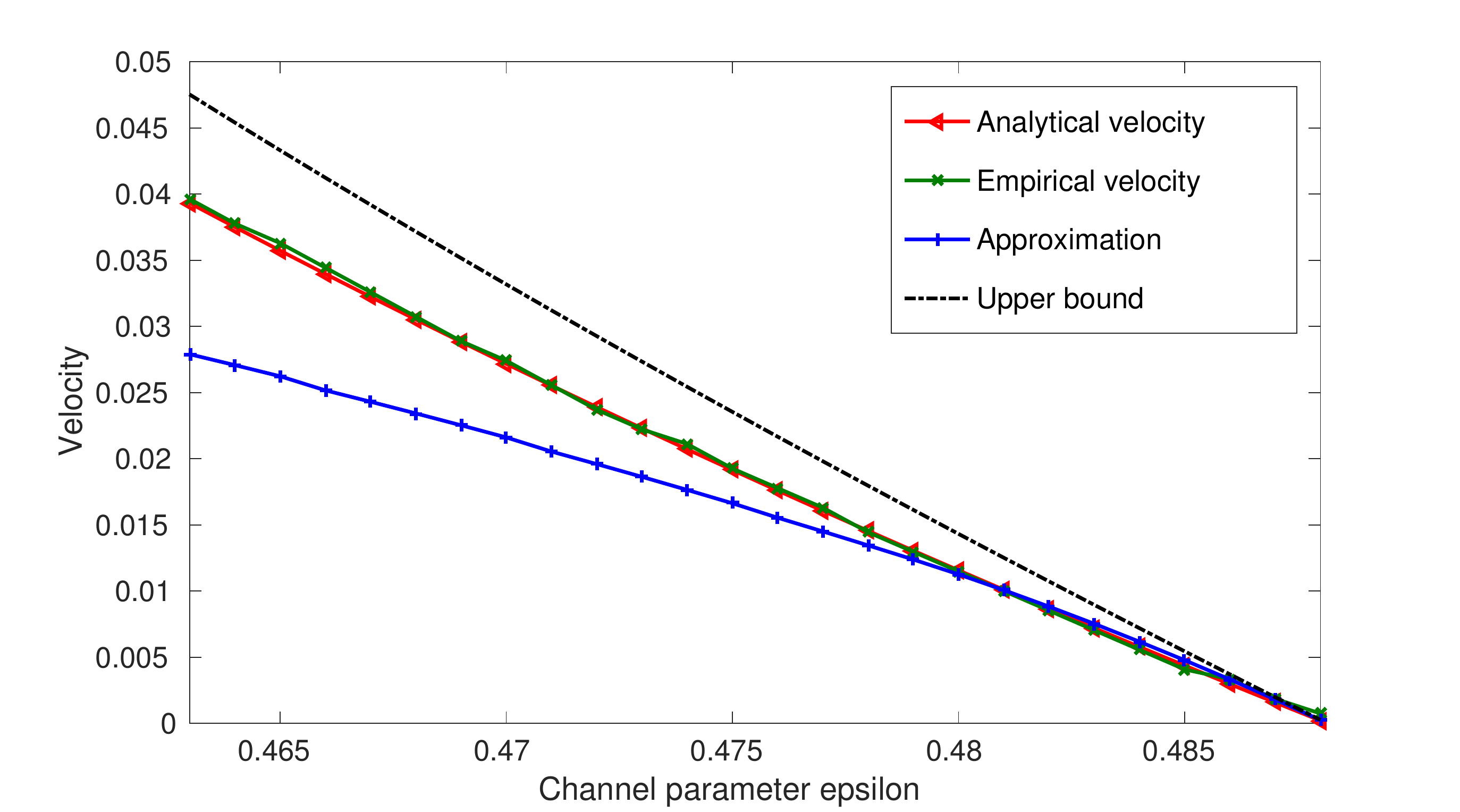}
\caption{We plot the normalized velocities $v_{\text{BEC}}$, $v_e$, $v_{a_2}$, and $v_{\text{B}}/\alpha$ (in the order of the legend) of the decoding profile for $0.463<\epsilon<\epsilon_{\text{\tiny MAP}}=0.4881$, for the $(3,6)$-regular code ensemble for $L=128$, $w=3$.}
\label{fig:velocityPlots36w3}
\end{figure}

Figure \ref{fig:velocityPlots36w3}  shows velocities (normalized by $w$) for the BEC for a $(3,6)$ coupled code with $L=128$ and $w=3$ in a range close to the MAP threshold.  The velocity decreases to zero as $\epsilon$ approaches $\epsilon_{\text{\tiny MAP}}$ and confirms that the soliton becomes ``static" at the MAP threshold (we point out that at $\epsilon_{\text{\tiny MAP}}$ the existence and unicity
up to translations of the static shape has been proven by displacement convexity techniques \cite{el2013displacement}, \cite{el2014analysis}). We see that the theoretical velocity $v_{\text{\tiny BEC}}$ and empirical velocity $v_e$ nicely match over a wide range of noise. 
The approximation $v_{a_2}$ works very well for $\epsilon$ close to $\epsilon_{\text{\tiny MAP}}$, and also with respect to $v_{\text{\tiny B}}$ with the added advantage that it does not require running DE. Similar findings are also illustrated 
in table \ref{table:VdiffE} for the coupled $(3,6)$-regular ensemble for $L=256$, $w=8$, and for different values of $\epsilon$.


Table \ref{table:vGauss} compares the theoretical and empirical velocities, $v_{\text{\tiny GA}}$ and $v_e$ for a 
BIAWGNC($\sigma_n^2$). 
The velocities are obtained for the $(3,6)$- and the $(4,8)$-regular ensembles for different values of $\psi^{-1}(H(c)) = 2/\sigma_n^2$ (twice the signal to noise ratio). 

\begin{table}[htb]
\caption{Normalized velocities of the wave on the LDPC($x^3,x^6$) on the BEC($\epsilon$) with $L=256$ for $w=8$
}
\label{table:VdiffE}
\begin{center}
\begin{tabular}{|c||c|c|c|c|}
\hline
{\bf $\epsilon$} & {\bf 0.455} & {\bf 0.465} & {\bf 0.475} & {\bf 0.485}\\
\hline
$v_{\text{BEC}}$ & 0.05754 & 0.03741 & 0.02004 & 0.00456 \\
\hline
$v_e$ & 0.05813  & 0.03750 & 0.02000 & 0.00468 \\
\hline
$v_{a_2}$ & 0.03470 & 0.02623 & 0.01663 & 0.00476 \\
\hline
$v_{\text{B}}/\alpha$ & 0.06108 & 0.03992 & 0.02149 & 0.00491 \\
\hline
\end{tabular}
\end{center}
\end{table}

\begin{table}[htb]
\caption{Normalized velocities of the waves on the LDPC($x^3,x^6$) and LDPC($x^4,x^8$) for the BIAWGNC($\sigma_n^2$) with $L=100$, $w=3$, within GA.
}
\label{table:vGauss}
\begin{center}
\begin{tabular}{|c||c|c|c|c|}
\hline
{\bf $2/\sigma_n^2$} & {\bf 2.33} & {\bf 2.35} & {\bf 2.38} & {\bf 2.40}\\
\hline
$v_{\text{\tiny GA}}$, $(3,6)$  & 0.0183 & 0.0222 & 0.0283 & 0.0325\\
\hline
$v_e$, $(3,6)$ & 0.0183 & 0.0233 & 0.0317 & 0.0375 \\
\hline
$v_{\text{\tiny GA}}$, $(4,8)$ & 0.0237 & 0.0258 & 0.0312 & 0.0381 \\
\hline
$v_e$, $(4,8)$ & 0.0217  & 0.0250 & 0.0308 & 0.0342  \\
\hline
\end{tabular}
\end{center}
\end{table}

\subsection{Scaling Law for Finite-Length Coupled Codes}\label{ssection:scalingLaw}
The authors in \cite{olmos2014scaling} propose a scaling law to predict the error probability of a finite-length spatially coupled $(\ell,r,L)$ ensemble when transmission takes place over the BEC. The derived scaling law depends on ``scaling parameters", one of which we will relate to the velocity of the decoding wave.
Note that the $(\ell,r,L)$ ensemble in \cite{olmos2014scaling} is slightly different than the one here but it is still of interest 
to discuss an application of the velocity formula to the scaling law.

Whenever a variable node is decoded, it is removed from the graph along with its edges. One way to track this peeling process is to analyze the evolution of the degree distribution of the residual graph across iterations, which serves as a sufficient statistic. This statistic can be described by a system of differential equations, whose solution determines the mean of the fraction of degree-one check nodes $\hat{r}_1$ and the variance (around this mean) at any time during the decoding process. As shown in \cite{olmos2014scaling} there exists a ``steady state phase" where the mean and the variance are constant, and during which one can observe the appearance of the decoding wave. (Note that here
we consider one-sided termination instead of two-sided termination in \cite{olmos2014scaling}, so the fraction $\hat{r}_1$ here is equal to half the fraction called $\hat{r}_1(*)$ in \cite{olmos2014scaling}).

Consider transmission over the BEC($\epsilon$) and let $\epsilon_{(\ell,r,L)}$ denote the BP threshold of the 
finite-size ensemble. We write the first-order Taylor expansion of $\hat{r}_1\big|_\epsilon$ around $\epsilon_{(\ell,r,L)}>\epsilon$ as
$\hat{r}_1\big|_\epsilon\approx\hat{r}_1\big|_{\epsilon_{(\ell,r,L)}}+\gamma\,\Delta\epsilon$
where $\Delta \epsilon=\epsilon_{(\ell,r,L)}-\epsilon$. Thus, since $\hat{r}_1\big|_{\epsilon_{(\ell,r,L)}}=0$ (by definition), then 
$\gamma\approx\hat{r}_1|_\epsilon/\Delta\epsilon$. This parameter $\gamma$ enters in the scaling law and is determined experimentally. 
Obviously it would be desirable to have a theoretical handle on $\gamma$.
It is argued in \cite{olmos2014scaling} that $\gamma\approx\bar{\gamma}$ where $\bar{\gamma}=x_{\text{\tiny BP}}\,v/\Delta\epsilon$ and where $v$ is the velocity of the decoding wave. 
We find that for the $(3,6)$  ensemble, $\gamma=2.155$, $\bar{\gamma}=1.960$; for the $(5,10)$ ensemble, $\gamma=2.095$, $\bar{\gamma}=1.733$; for the $(4,12)$ ensemble, $\gamma=2.140$, $\bar{\gamma}=1.778$, to list a few examples. The differences might mostly be related to the difference in ensembles considered here and in 
\cite{olmos2014scaling}, and also to the relatively large value of $\Delta\epsilon=0.04$ (chosen in \cite{olmos2014scaling} due to stability issues in numerical simulations).


\section*{Acknowledgment}
R. E. thanks Andrei Giurgiu for finding the bug in her code. We also thank R{\"u}diger Urbanke for 
discussions.

\bibliographystyle{IEEEtran}
\bibliography{BIBfile}

\end{document}